\documentclass[a4paper,fleqn,usenatbib,useAMS]{mnras}

\usepackage{newtxtext,newtxmath}
\usepackage{graphicx}	
\usepackage{amsmath}	
\usepackage{amssymb}	
\usepackage{gensymb}
\usepackage{supertabular}
\usepackage[export]{adjustbox}

\newcommand{\kms}{\hbox{km s$^{-1}$}}
\newcommand{\vsini}{\hbox{$v \sin i$}}

\title[Doppler imaging of RS CVn]{The first Doppler imaging of the active binary prototype RS Canum Venaticorum}
\author[Y.~Xiang et al,]{Yue Xiang,$^{1}$$^{2}$\thanks{E-mails: shenghonggu@ynao.ac.cn(SG); xy@ynao.ac.cn (YX)} Shenghong Gu,$^{1}$$^{2}$$^{3}$$^{\star}$ U. Wolter,$^{4}$ J. H. M. M. Schmitt,$^{4}$ \and A.~Collier~Cameron,$^{5}$ J.~R.~Barnes,$^{6}$ M. Mittag,$^{4}$ V. Perdelwitz$^{4}$ and S. Kohl$^{4}$\\
$^{1}$Yunnan Observatories, Chinese Academy of Sciences, Kunming 650216, China\\
$^{2}$Key Laboratory for the Structure and Evolution of Celestial Objects, Chinese Academy of Sciences, Kunming 650216, China\\
$^{3}$University of Chinese Academy of Sciences, Beijing 100049, China\\
$^{4}$Hamburger Sternwarte, Universit\"{a}t Hamburg, Hamburg 21029, Germany\\
$^{5}$School of Physics and Astronomy, University of St Andrews, Fife KY16 9SS, UK\\
$^{6}$Department of Physical Sciences, The Open University, Walton Hall, Milton Keynes MK7 6AA, UK\\
}

\begin{document}

\label{firstpage}
\pagerange{\pageref{firstpage}--\pageref{lastpage}}
\maketitle

\begin{abstract}
We present the first Doppler images of the prototypical active binary star RS CVn, derived from high-resolution spectra observed in 2004, 2016 and 2017, using three different telescopes and observing sites. We apply the least-squares deconvolution technique to all observed spectra to obtain high signal-to-noise line profiles, which are used to derive the surface images of the active K-type component. Our images show a complex spot pattern on the K star, distributed widely in longitude. All starspots revealed by our Doppler images are located below a latitude of about 70{\degree}. In accordance with previous light-curve modeling studies, we find no indication of a polar spot on the K star. Using Doppler images derived from two consecutive rotational cycles, we estimate a surface differential rotation rate of $\Delta\Omega = -0.039 \pm 0.003 ~rad~d^{-1}$ and $\alpha = \Delta\Omega/\Omega_{eq} = -0.030 \pm 0.002$ for the K star. Given the limited phase coverage during those two rotations, the uncertainty of our differential rotation estimate is presumably higher.
\end{abstract}


\section{Introduction}

RS CVn-type stars, as defined by \citet{hall1976}, are a class of close binary systems consisting of a chromospherically active subgiant component, which exhibits the brightness variations caused by large cool spots. The prototype star, RS CVn, is an eclipsing binary system composed by an F5 main-sequence and a K2 subgiant stars\citep{reg1990,rod2001}, with an orbital period of 4.797695 d \citep{eaton1993}. \citet{pop1988} estimated a different spectral type of F4 + G9IV for RS CVn, while \citet{str1990} determined those of two components to be F6IV + G8IV. \citet{bar1994} determined the position of the two components on the H-R diagram, which indicated an age of 2.5 Gyr for RS CVn. The mass transfer between two components of RS CVn-type binaries is not relevant for their phenomenology making them more suitable to study magnetic activity than systems with mass transfer such as Algols.

The distortions in the light curves of RS CVn are attributed to the non-uniform distributed cool spots \citep{eaton1979,kang1989}. \citet{rod1995} analysed the long-term sequence photometric data of RS CVn and estimated a period of 19.9 yr for its starspot activity. They also found a spot migration rate of 0.1{\degree} per day during 1963--1984 and a rate of 0.34{\degree} per day during 1988--1993, with respect to the rotating frame of the K star. They inferred a solar-like surface differential rotation for the K-type component and the shear rate is about 5--20 percent of the value of the Sun. Their O-C diagram of the epochs of the primary minima indicates that the orbital period of RS CVn is changing on a time scale of an order of 100 yr, and the O-C variations are of the order of almost 0.3 d. Such period variations in close binary systems have been proposed to be caused by strong magnetic activity \citep{app1992}. \citet{rod2001} further determined accurate system parameters using a long-term sequence of the light curves of RS CVn, taking into account the light curve distortions caused by starspots.

Through the analysis of the multi-color photometry data for RS CVn, \citet{aar2005} found that the hot faculae surrounding cool starspots on the surface of the cooler component were necessary to explain the observed colour variation. \citet{mes2008} inferred that RS CVn's activity only takes place in the K-type component, and the short-term color variations are dominated by faculae, whereas the long-term colour variation can be partly caused by the F-type component that makes the system appear bluer when the K component becomes fainter owing to its variable starspot coverage. From the combination of the photometric and spectroscopic observations of RS CVn, \citet{eaton1993} found that several moderately sized spots on the surface of the cooler component are needed to fit the observed data. They did not find any evidence for large polar spots.

RS CVn also exhibits signs of magnetic activity in the chromospheric indicators \citep{fer1994,mon1996}, as well as coronal emission. Its X-ray luminosity of $log(L_{X})=31.33~erg~s^{-1}$ \citep{dra1992}, in combination with the bolometric luminosity derived by Gaia \citep{gaia2016,gaia2018}, yields $log(L_{X}/L_{bol})=-3.27$, placing the system close to the saturation limit for late-type main-sequence stars of $log(L_{X}/L_{bol})=-3$ \citep{piz2003}.

Up to now, there are only photometric and limited spectroscopic studies on the starspot activity of this prototype star, and no Doppler image derived for it. The Doppler imaging technique can offer a better constraint on spot latitudes. In order to investigate the spot activities of active binary stars, we continued to monitor a set of RS CVn-type binary systems\citep{gu2003,xiang2014,xiang2015,xiang2016}. In this study, we performed high-resolution spectroscopic observations on the active binary prototype RS CVn using three telescopes located at different observing sites. We applied the Doppler imaging code to derive the first detailed spot maps of the K-type component of RS CVn for 2004 February, 2016 January, 2017 April and 2017 November--December.

In section 2, we describe the spectroscopy observations and data reduction. In section 3, we offer the results of Doppler imaging of K-type component of RS CVn. We discuss the distribution of starspots on RS CVn and its surface differential rotation in Section 4. In Section 5, we summarise our results for RS CVn.

\section{Observations and data reduction}

The high-resolution spectroscopic observations of RS CVn were carried out at three different observing sites, in 2004 February, 2016 January, 2017 April and 2017 November--December. A Coud\'{e} echelle spectrograph (CES; \citealt{zhao2001}) with a 1024 $\times$ 1024 pixel Tektronix CCD detector mounted on the 2.16m telescope at Xinglong station of National Astronomical Observatories of China, was used to collect spectra of RS CVn on 2004 February 3--9. Its resolution power is R = 37 000 and the spectral coverage is 5500--9000{\AA}. The exposure time varied in a range from 1200s to 3600s, depending on the weather. On 2016 January 22--31 and 2017 November 29--December 11, a new fibre-fed, high-resolution spectrograph (HRS) with a 4096 $\times$ 4096 pixel EEV CCD detector installed on the 2.16m telescope at Xinglong station was used to acquire spectral data. It has a resolution power of R = 48 000 and covers from 3900 to 10000 {\AA}. The exposure time of all these observations was fixed to 1800s.

A joint observation campaign was carried out by using the 1.2m robotic spectroscopy telescope TIGRE \citep{sch2014} at La Luz Observatory, Guanajuato, Mexico and the 1m telescope of Shandong University at Weihai on 2017 April 13--21. The TIGRE telescope was equipped with a fiber-fed echelle spectrograph HEROS which has a resolution power of R $\approx$ 20 000 and a spectral coverage of 3800--8800 {\AA} with a small gap of 100 {\AA} around 5800 {\AA}. The fiber-fed spectrograph mounted on the 1m telescope is the same type as the one on the 2.16m telescope used in 2016 and 2017. Due to the bad weather, the 1m telescope only collected six spectra with sufficient signal-to-noise ratio (SNR), and others are removed to avoid artefacts. We summarised the observations in Table \ref{tab:obs}, and listed them in detail in Appendix A, which is only available online, including UT date, HJD, orbital phase, exposure time and the peak SNR of each observed spectrum.

\begin{table}
 \caption{Summary of observations.}
 \label{tab:obs}
 \begin{tabular}{lccc}
 \hline
 Date & Instrument & Resolution & No. of spectra\\
 \hline
 2004 Feb 3--9       & 2.16m/CES & 37 000 & 8\\
 2016 Jan 23--31     & 2.16m/HRS & 48 000 & 26\\
 2017 Apr 14--21     & TIGRE/HEROS & 20 000 & 46\\
 2017 Apr 18         & 1m/HRS    & 48 000 & 6\\
 2017 Nov 28--Dec 11 & 2.16m/HRS & 48 000 & 29\\
 \hline
 \end{tabular}\\
\end{table}

The spectral data collected with the TIGRE telescope were reduced with the TIGRE data reduction pipeline \citep{mit2010}. The data obtained from 2.16m and 1m telescopes were reduced using the IRAF\footnote{IRAF is distributed by the National Optical Astronomy Observatory, which is operated by the Association of Universities for Research in Astronomy (AURA) under cooperative agreement with the National Science Foundation.} package in a standard way, which included image trimming, bias subtraction, flat-field dividing, scatter light subtraction, cosmic-ray removal, 1D spectrum extraction, wavelength calibration and continuum fitting. The wavelength calibration was performed by using the comparison spectra of the ThAr lamp taken at each night.

\section{Least-Squares Deconvolution}

In order to enhance the SNR of the stellar line profiles, we applied the Least-Squares Deconvolution (LSD; \citealt{don1997}) technique to all observed spectra. The LSD technique combines all available observed photospheric lines to derive an average line profile with much higher SNR. The line list, including the central wavelength and depth of spectral lines, was derived from the Vienna Atomic Line Database (VALD; \citealt{kup1999}). In our case, we used the standard LSD technique and only used the line list of the K component in the computation. \citet{tka2013} showed that the using of the line list of only one component results in the incorrect depth of the profile of the other component but has little effect on the shapes and the radial velocities of the line profiles of two stars. We used all available photospheric lines except for those around strong telluric and stellar chromospheric lines to avoid their effects. The SNR of each resulting LSD profile is also listed in Appendix A. The SNR was typically improved by a factor of about 15. For each observing run, we also derived the telluric line profiles from the observed spectra and cross-correlated them to calculate and correct the small instrumental shifts (smaller than 0.5 \kms\ for our data sets) which would otherwise introduce errors in the radial velocities and thus improve the Doppler imaging and orbital solutions. This correction method was developed by \citet{cam1999} and can achieve a precision better than 0.1 {\kms} \citep{don2003}.

Our imaging code employs the two-temperature model, which treats the stellar surface as a composition of only two temperature components, the hot photosphere and cool spot, and uses the spot filling factor $f$ to represent the fractional spottedness of each image pixel \citep{cam1994}. Thus the spectra of the template stars (Table \ref{tab:template}) for the photospheres and starspots were also deconvolved in the same manner as that for the spectra of RS CVn to construct the lookup tables, which contain the local intensity profiles of two temperature components at different limb angle on the stellar surface. We obtained the linear limb-darkening coefficients of UBVRI passbands from \citet{cla2012,cla2013}, for the effective temperatures of the photospheres and spots of the F main-sequence and K subgiant stars. The spot temperature of 3500 K was chosen according to the values of similar systems. Given the fact that the limb-darkening coefficient is almost a linear function of the wavelength, for each temperature component of each star we used a linear interpolation to derive the value at the centroidal wavelength of 6170 \AA, which was calculated from the line list \citep{bar1998}. The results are 0.84 and 0.77 for the starspot and photosphere of the K star, 0.76 and 0.55 for those of the F star, respectively. Then we calculated the local intensity at 30 limb angles to produce the lookup tables.

\begin{table}
 \caption{Template stars for the photoshperes and spots of two components for each observing site.}
 \label{tab:template}
 \begin{tabular}{lccc}
 \hline
 Spectra type & TIGRE & Weihai 1m & Xinglong 2.16m\\
 \hline
 F6V  & HD 216385 & HR 3262 & HR 3262 \\
 K2IV & HR 5227   & HR 5227 & HR 8088 \\
 M0IV (spot) & HR 4920 & HR 4920 & HR 4920 \\
 \hline
 \end{tabular}
\end{table}

\section{Doppler imaging}

\subsection{System parameters}

The Doppler imaging technique requires accurate stellar parameters, such as the projected rotational velocity ({\vsini}), inclination and rotational period, to derive reliable surface maps and to prevent producing artefacts \citep{cam1994}, especially for eclipsing binary systems \citep{vin1993}. It has been demonstrated that the Doppler imaging code can also be used to determine the stellar parameters of single and binary stars (e.g., \citealt{bar1998,bar2004}). Fine-tuning stellar parameters can be achieved by performing a fixed number of the maximum entropy iterations with various combinations of stellar parameters and then finding the best-fit values which leads to a minimum $\chi^{2}$. This method can overcome the effect of starspot distortions on the parameter determinations \citep{bar2005}.

In this work, we performed the $\chi^{2}$ minimization method to estimate best stellar parameters for Doppler imaging of RS CVn. Since the orbital elements of RS CVn have been widely studied (e.g., \citealt{cat1974,pop1988,eaton1993}), we adopted the values of the inclination (i), the orbital period (P$_{orbit}$) and conjunction time (T$_{0}$) derived by \citet{eaton1993} and the Albedo coefficients of the two stars in the paper of \citet{rod2001} as the fixed parameters which did not change in the procedure. We used the values of the mass ratio (q), the radial velocity amplitudes of the two stars (K) derived by \citet{eaton1993} as the initial guess, and performed a grid search within a small range around them to search for the best-fit values. With a small systematic orbital phase offset in a range of 0.0007--0.0019 for each observing run, which takes the long-term orbital period variation into account, we found that the orbital ephemeris taken from \citet{eaton1993} is sufficient for Doppler imaging. We show the deviation of orbital phase and the corresponding conjunction time between the values calculated from our data sets and those from the ephemeris of \citet{eaton1993} in Table \ref{tab:t0}. We list the adopted values of the stellar parameters for imaging RS CVn in Table \ref{tab:par}.

\begin{table}
 \caption{Deviation of orbital phase and the corresponding $\Delta$T$_{0}$ between the values calculated from our data sets and those from the ephemeris of \citet{eaton1993}.}
 \label{tab:t0}
 \begin{tabular}{lcc}
 \hline
 Epoch (mean HJD) & $\Delta\phi$ & $\Delta$T$_{0}$\\
 +2450000    & O$-$C       & O$-$C (day)\\
 \hline
 3042.5 & -0.0011 & 0.0052 \\
 7415.4 & -0.0019 & 0.0091 \\
 7861.9 & -0.0007 & 0.0034 \\
 8094.0 & -0.0013 & 0.0062 \\
 \hline
 \end{tabular}
\end{table}

\begin{table}
 \caption{Adopted stellar parameters of RS CVn for Doppler imaging.}
 \label{tab:par}
 \begin{tabular}{lcc}
 \hline
 Parameter & Value & Ref.\\
 \hline
 $q=M_{K}/M_{F}$ & $1.07 \pm 0.01$ & DoTS\\
 $K_{F}$ (km s$^{-1}$) & $90.2 \pm 0.1$ & DoTS\\
 $K_{K}$ (km s$^{-1}$) & 84.3 & $K_{F}$ and q\\
 $i$ (\degree) & 85.55 & a\\
 $T_{0}$ (HJD) & 2448379.1993 & a\\
 $P_{orbit}$ (d)  & 4.797695 & a\\
 $\vsini_{F}$ (km s$^{-1}$) & $12.4 \pm 0.5$ & DoTS\\
 $\vsini_{K}$ (km s$^{-1}$) & $44.9 \pm 1.0$ & DoTS\\
 $R_{F}$ ($R_{\odot}$) & 2.1 & DoTS\\
 $R_{K}$ ($R_{\odot}$) & 4.3 & DoTS\\
 $P_{rot, F} (d)$ & 8.542 & \vsini, $i$ and $R_{F}$ \\
 $T_{eff, F}$ (K)& 6800 & a\\
 $T_{eff, K}$ (K)& 4580 & a\\
 Albedo$_{F}$ & 1.0 & b\\
 Albedo$_{K}$ & 0.3 & b \\
 \hline
 \end{tabular}\\
  References: a. \citet{eaton1993}; b. \citet{rod2001}.\\
\end{table}

\begin{figure}
\centering
\includegraphics[width=0.95\linewidth]{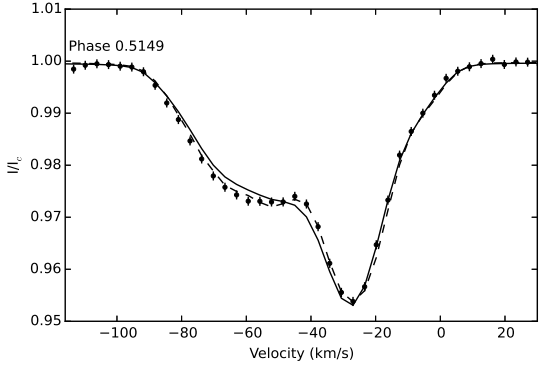}
\caption{A comparison of the fits to the LSD profile observed at eclipsing phase 0.5149 with 2.16m/HRS, assuming synchronous (solid line) or non-synchronous rotation (dashed line) of the F component.}
\label{fig:nodr}
\end{figure}

In the imaging process, we noticed that we can not obtain a good fit with the assumption of the synchronous rotation. The F-type component seems to have a significantly smaller rotational speed, similar to another RS CVn-type system SZ Psc \citep{xiang2016}. As an example, Fig. \ref{fig:nodr} shows the fit to the observed LSD profile at phase 0.5149, where the F star was eclipsing the K star. The misfit is obvious if our imaging code treats it as a synchronous binary system. Given a $\vsini$ of 12.4 {\kms} derived from line profile broadening, the radius of the F star is only 1.2 R$_{\odot}$ assuming it is tidally locked. By contrast, \citet{eaton1993} derived a stellar radius of about 2.0 R$_{\odot}$ for the F star from the light curve modelling. \citet{str1990} and \citet{eaton1993} found that the difference between the widths of the observed and calculated profiles for the F star can be attributed to the non-synchronous rotation.

To deal with it, we have a minor modification on the Doppler imaging code to take into account the non-synchronous rotation of the F-type component. This is achieved by recalculating the positions of the stellar grids and the velocity of pixels in the view plane when integrating flux of the F star, according to its rotational rate and stellar radius. The detailed modification method was described in the previous paper \citep{xiang2016}. As a result, we found that the F star rotates about 1.8 times slower than expected on the basis of synchronous rotation. The projected rotational velocity of the F star is $\vsini = 12.4~\kms$, whereas the synchronous rotation velocity derived from the orbital period and the stellar radius is $\vsini = 22.3~\kms$. Our result is consistent with that of \citet{str1990}. Their estimates of the rotation rates of the F and K stars are $11 \pm 2~\kms$ and $42 \pm 3~\kms$, respectively.

\subsection{Spot images}

\begin{figure}
\centering
\includegraphics[width=0.75\linewidth]{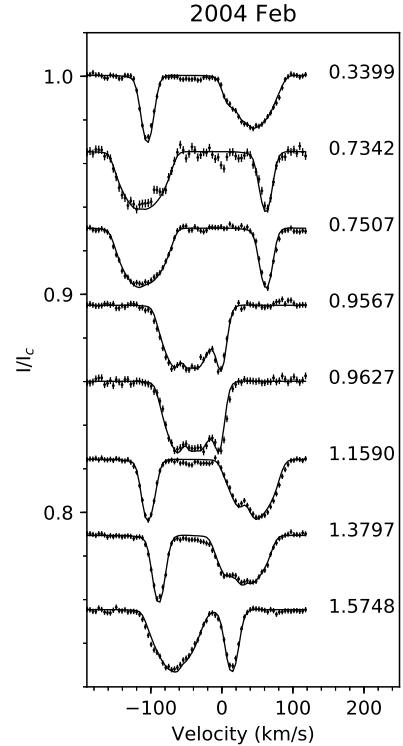}
\caption{Doppler imaging line profile fits (lines) and data set (dots with 1 $\sigma$ error bars) collected by 2.16m/CES in 2004 February. The rotational cycles are marked beside each profile. The reduced $\chi^{2}$ = 1.41.}
\label{fig:f0402}
\end{figure}

\begin{figure*}
\centering
\includegraphics[width=0.95\textwidth]{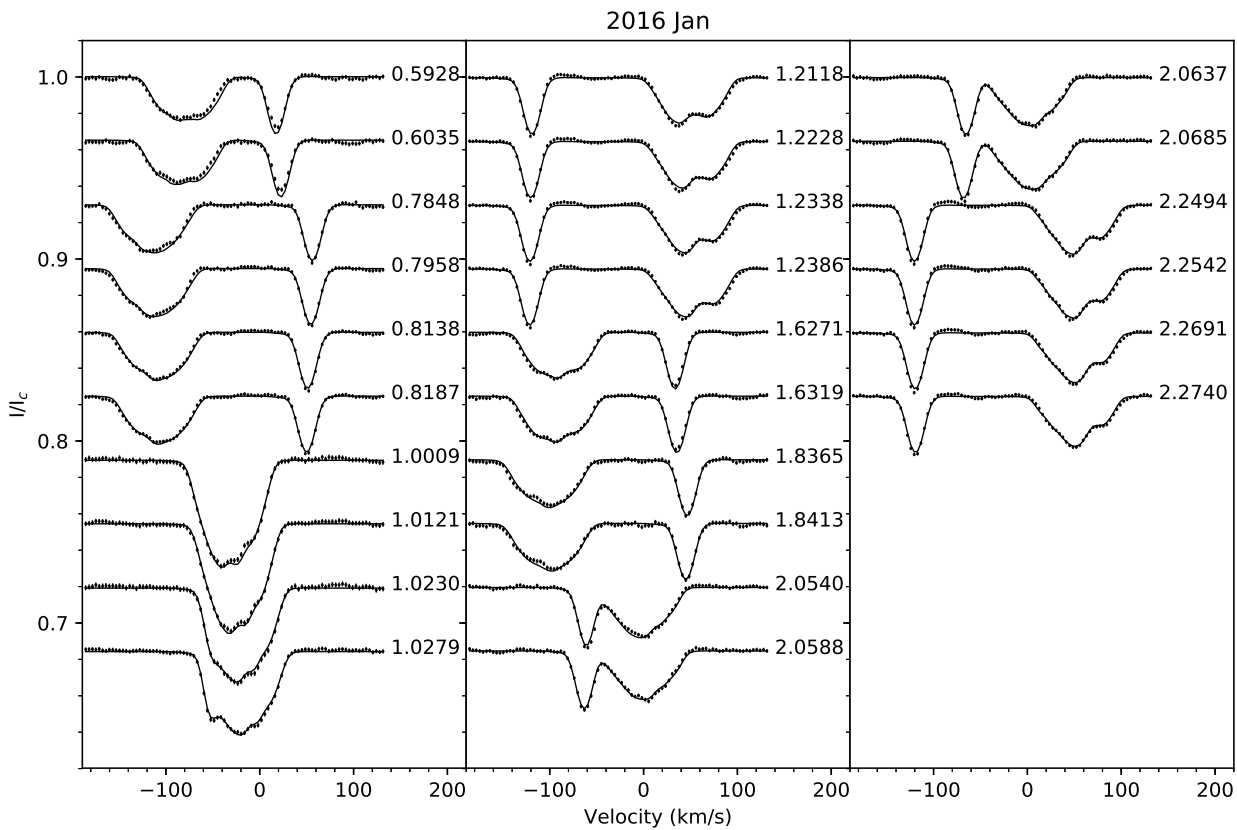}
\caption{Same as Fig. \ref{fig:f0402}, but for the data set collected by 2.16m/HRS in 2016 January. The reduced $\chi^{2}$ = 1.29.}
\label{fig:f1601}
\end{figure*}

\begin{figure*}
\centering
\includegraphics[width=0.95\textwidth]{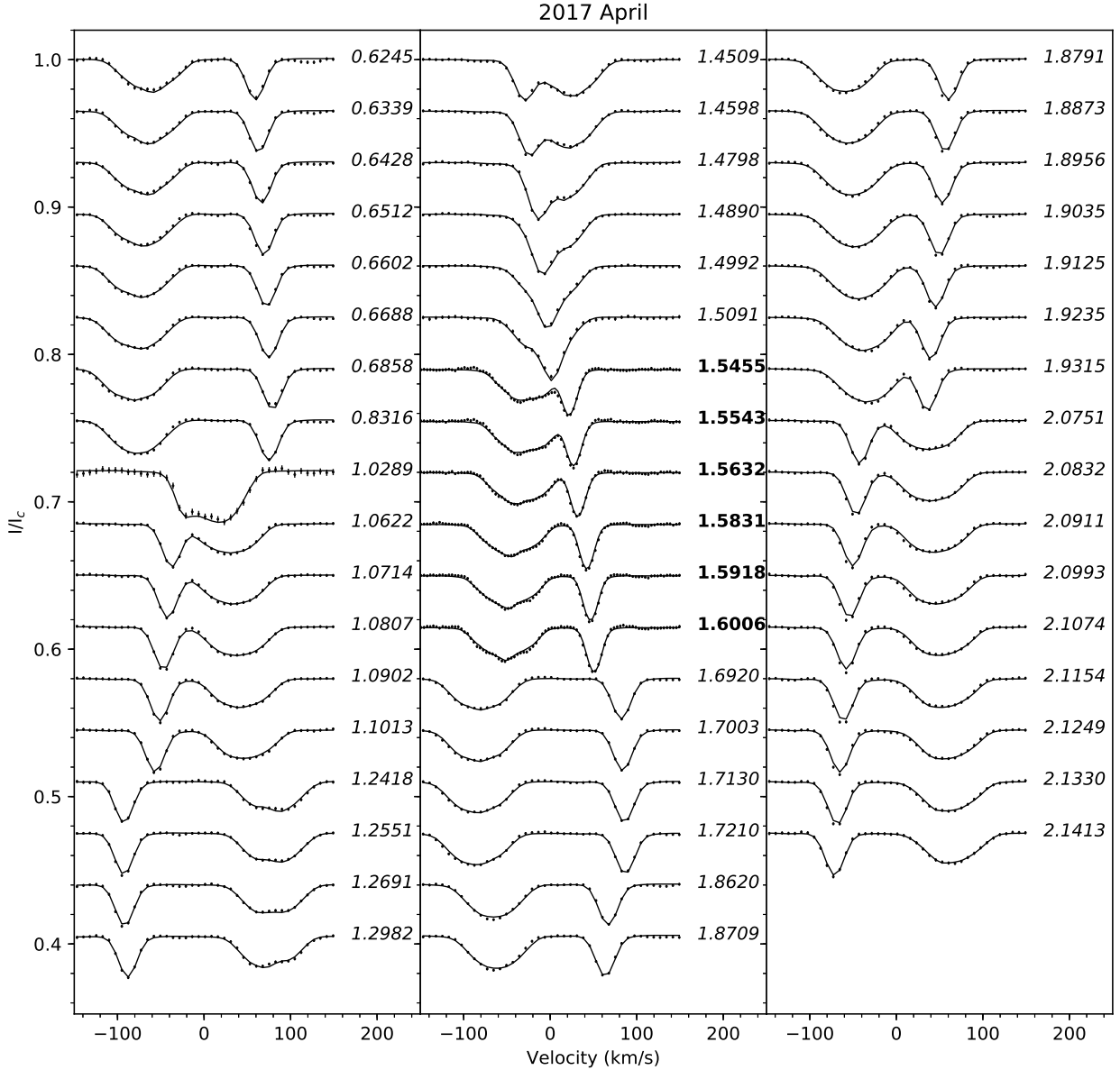}
\caption{Same as Fig. \ref{fig:f0402}, but for the data sets collected by TIGRE/HEROS (phase in italic) and 1m/HRS (phase in bold) in 2017 April. The reduced $\chi^{2}$ = 1.07.}
\label{fig:f1704}
\end{figure*}

\begin{figure*}
\centering
\includegraphics[width=0.95\textwidth]{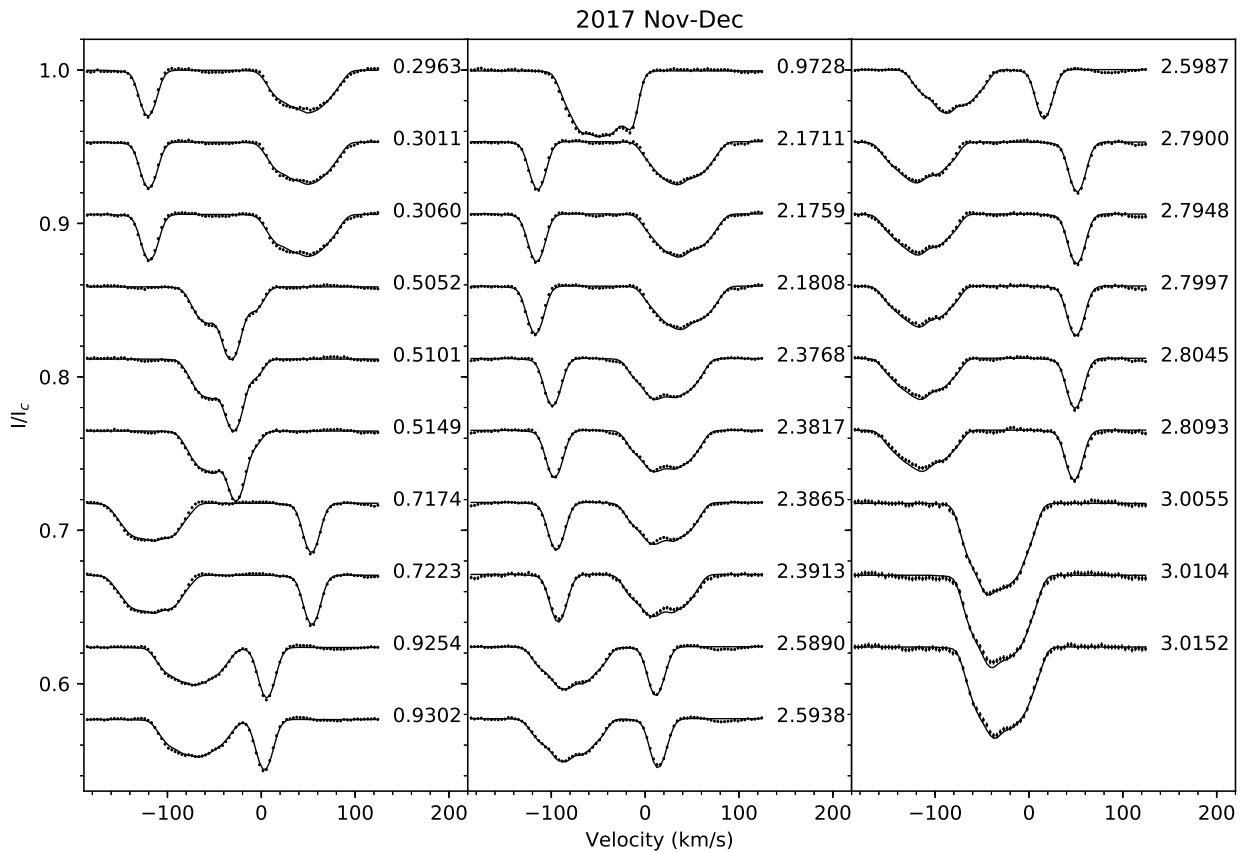}
\caption{Same as Fig. \ref{fig:f0402}, but for the data set collected by 2.16m/HRS in 2017 November--December. The reduced $\chi^{2}$ = 1.26.}
\label{fig:f1711}
\end{figure*}

\begin{figure*}
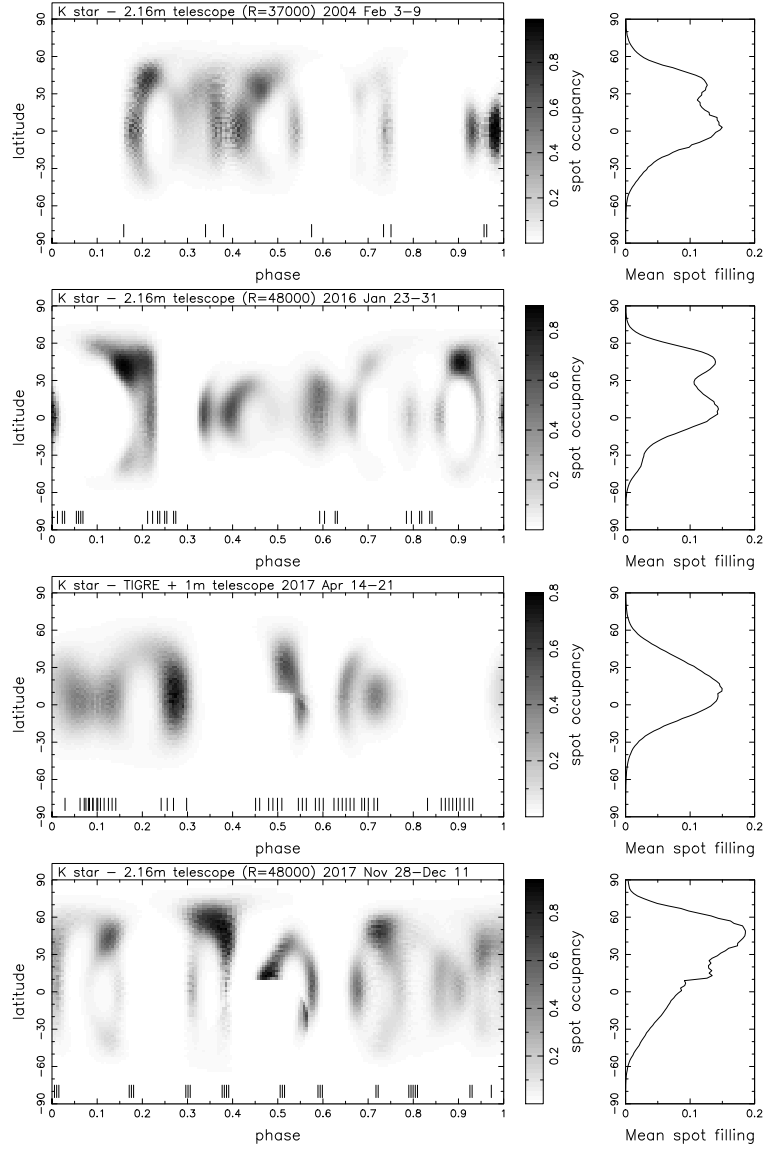

\centering
\includegraphics[bb = 72 18 360 776, angle=270, width=0.56\textwidth]{i0402.eps}
\includegraphics[bb = 72 18 360 776, angle=270, width=0.56\textwidth]{i1601.eps}
\includegraphics[bb = 72 18 360 776, angle=270, width=0.56\textwidth]{i1704.eps}
\includegraphics[bb = 72 18 360 776, angle=270, width=0.56\textwidth]{i1711.eps}
\caption{Doppler images of K star for 2004 February, 2016 January, 2017 April, 2017 November--December. The observed phases are marked by the vertical ticks. The mean spot filling factor vs latitude is plotted beside each image. The equatorial spots at phase 0.5 show sharp edges due to the passage of the F star.}
\label{fig:image}
\end{figure*}

We used the imaging code Doppler Tomography of Star (DoTS; \citealt{cam1992}; \citealt{cam1997}) to implement the maximum entropy regularized iterations to all of the data sets. Fig. \ref{fig:f0402}--\ref{fig:f1711} show the maximum entropy fits to the observed LSD profiles of the data sets in 2004 February, 2016 January, 2017 April and 2017 November--Decembers. The reduced $\chi^{2}$ for each data sets is shown in the caption of each figure. We did not achieve a reduced $\chi^{2}$ of 1.0, due to the underestimated errors in the LSD computation \citep{don1997,bar2005} and the maximum entropy regularization. Fig. \ref{fig:image} shows the reconstructed images of the K-type component of RS CVn, derived from these four data sets. The mean spot filling factor as a function of latitude is also plotted for each image. Given the \vsini\ of the K star of RS CVn and the resolution power of the spectrograph, the data sets collected by 2.16m/CES, 2.16m/HRS, 1m/HRS and TIGRE/HEROS respectively offered about 12, 15, 15, 6 resolution elements across the stellar disk, which translate into the longitudinal resolution of about 15\degree, 12\degree, 12\degree\ and 30\degree, respectively, for the Doppler imaging \citep{bor2019}.

In our spot maps, phase 0.5 (longitude 180\degree) on the K-type component faces the F-type component. One should notice an obvious spurious feature in all of the Doppler images that the low latitude spots are smeared and elongated vertically. This is due to the poor latitude resolution of the Doppler imaging technique for the low-latitude features. Hence the shape of low-latitude spots should not be over interpreted. The mirroring effect is strong for the high inclination star like RS CVn, but can be broken by the eclipse \citep{vin1993}. The spots in the images of the K star of 2017 April and November-December show sharp edges around phase 0.5 due to the passage of the F star.

The surface image of 2004 February shows connected structures between phases 0.2 and 0.6, composed by several spots. The features extend from latitude 0{\degree} to 60{\degree}. Two connected spots appear at the stellar equator around phase 0.95. The image of 2016 January shows several spot groups at phases 0, 0.2, 0.4, 0.6, 0.9. A strong spot is located at latitude 45{\degree} and phase 0.2 with appendages extending to the stellar equator.

The 2017 April data set has the best phase coverage among four observing runs. The image shows strong spots at the equator around phases 0.3 and 0.5, and a spot at latitude 30{\degree} and phase 0.7. These spots are also appended by weak spot features. The image derived for 2017 November--December, which is about half a year apart from the 2017 April observing run, shows a similar spot pattern. But the latitudinal concentration of the spot activity changed from 10{\degree} to 50{\degree}, as seen in the mean spot filling factor vs latitude plot.  The spot at phase 0.3 in 2017 April disappeared in 2017 November--December, while a strong spot emerged at phase 0.4 latitude 60{\degree}. The region around phase 0.5 was still active, and the spot group around phase 0.7 seemed to retain, but its position and distribution changed.

\begin{figure*}
\centering
\includegraphics[angle=270, width=0.35\textwidth]{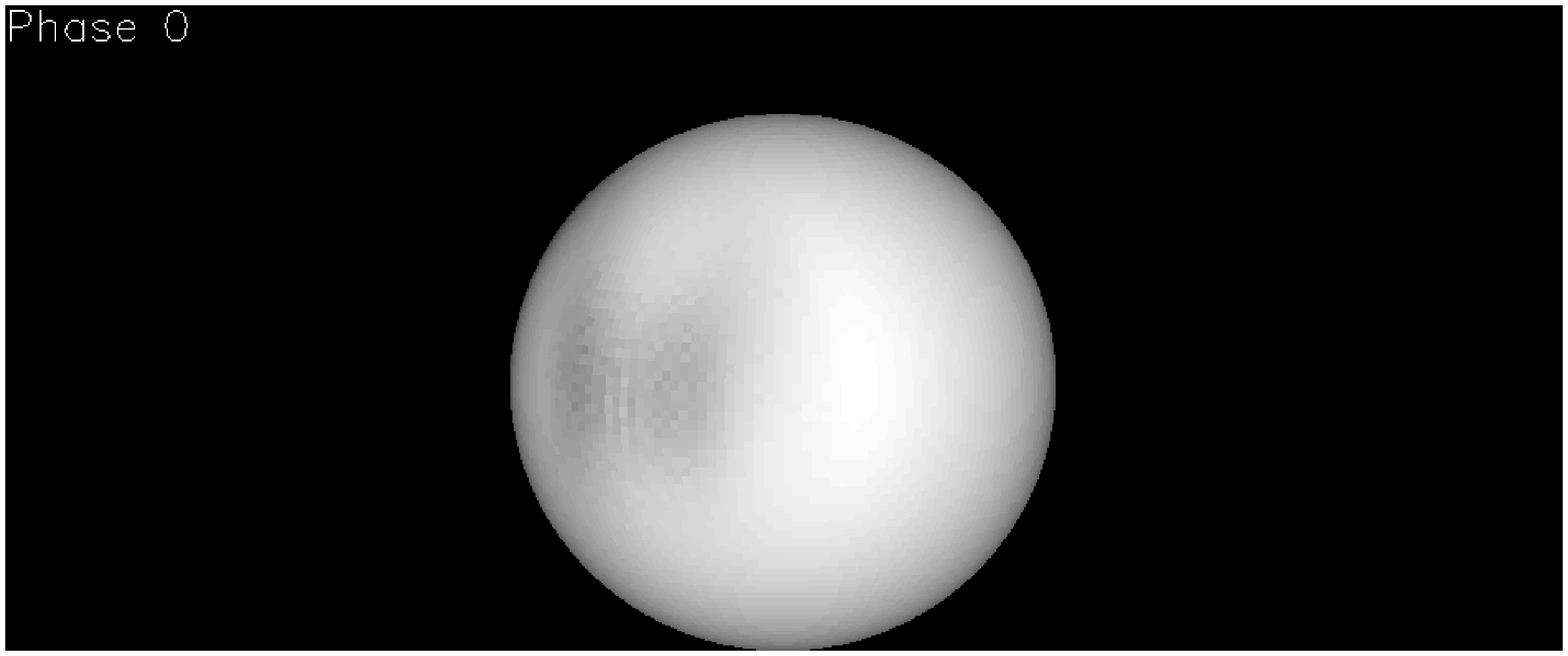}
\includegraphics[angle=270, width=0.35\textwidth]{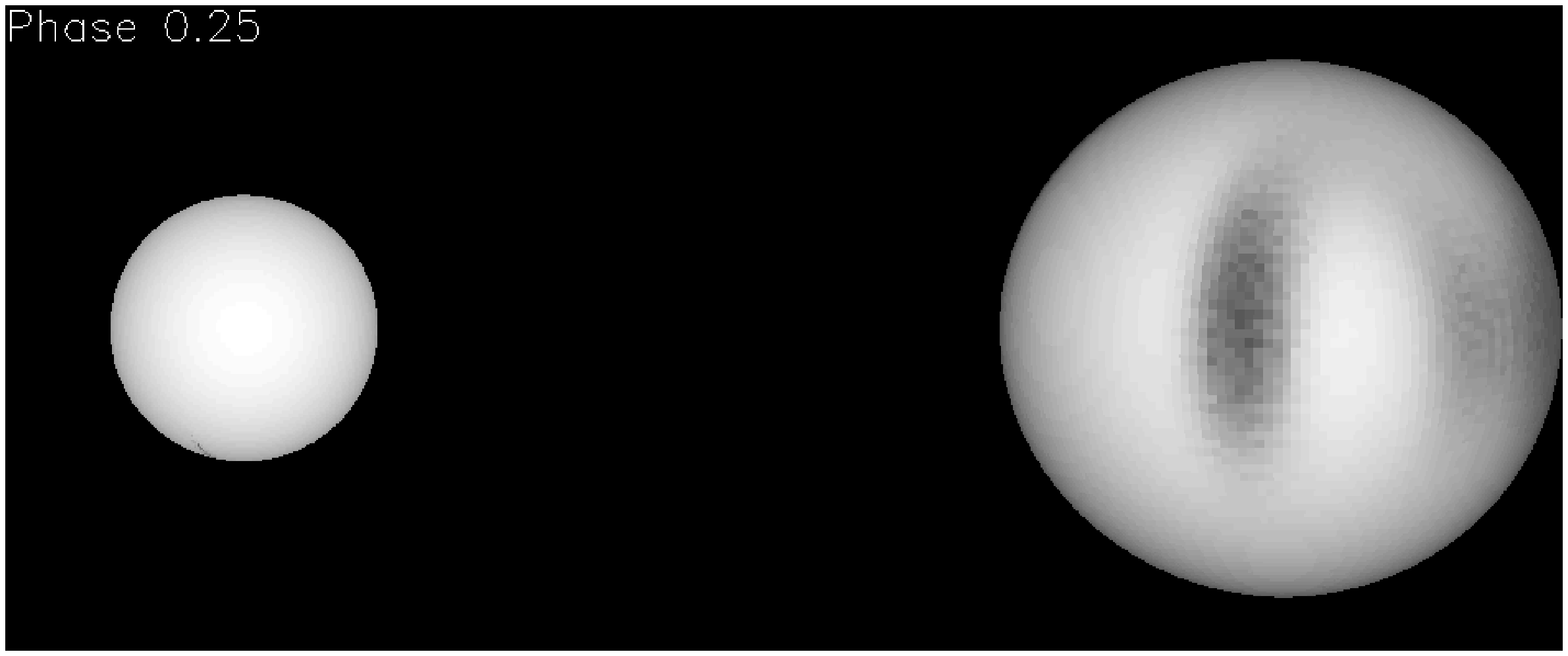}
\includegraphics[angle=270, width=0.35\textwidth]{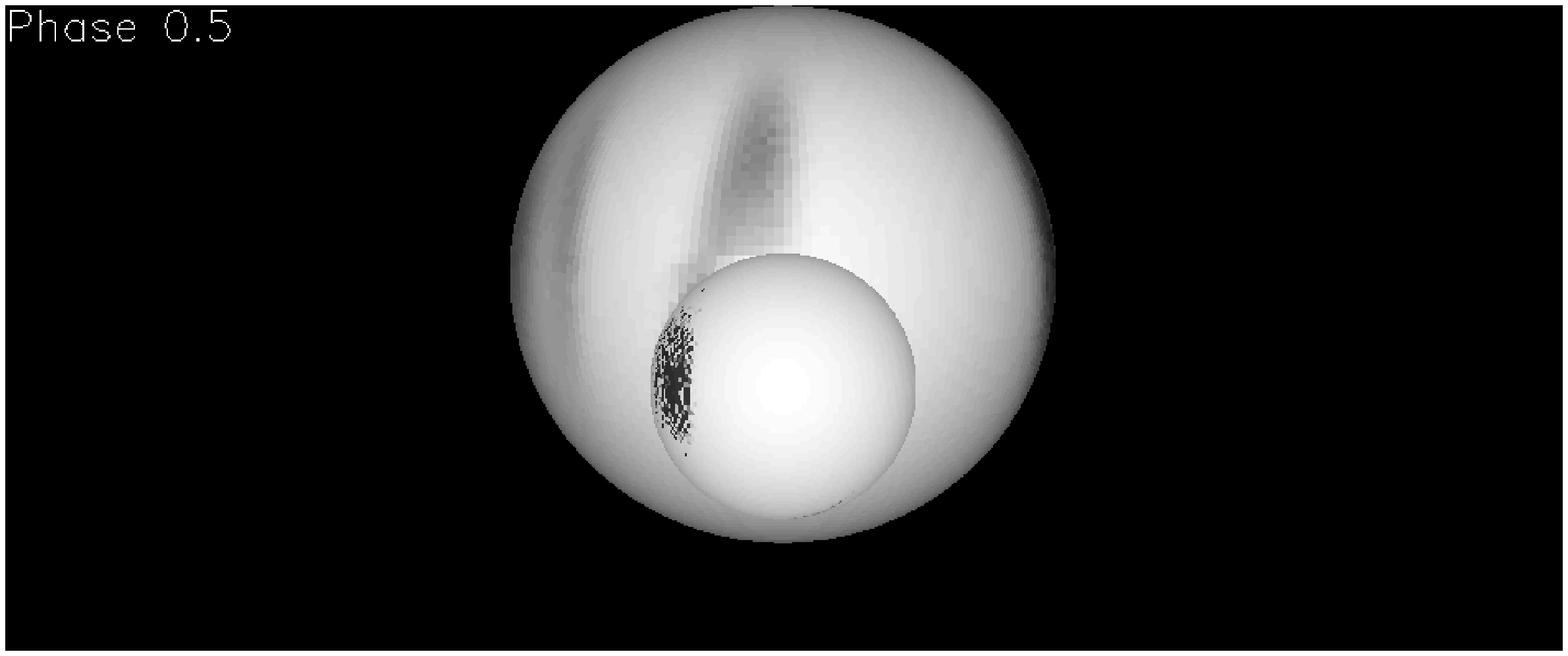}
\includegraphics[angle=270, width=0.35\textwidth]{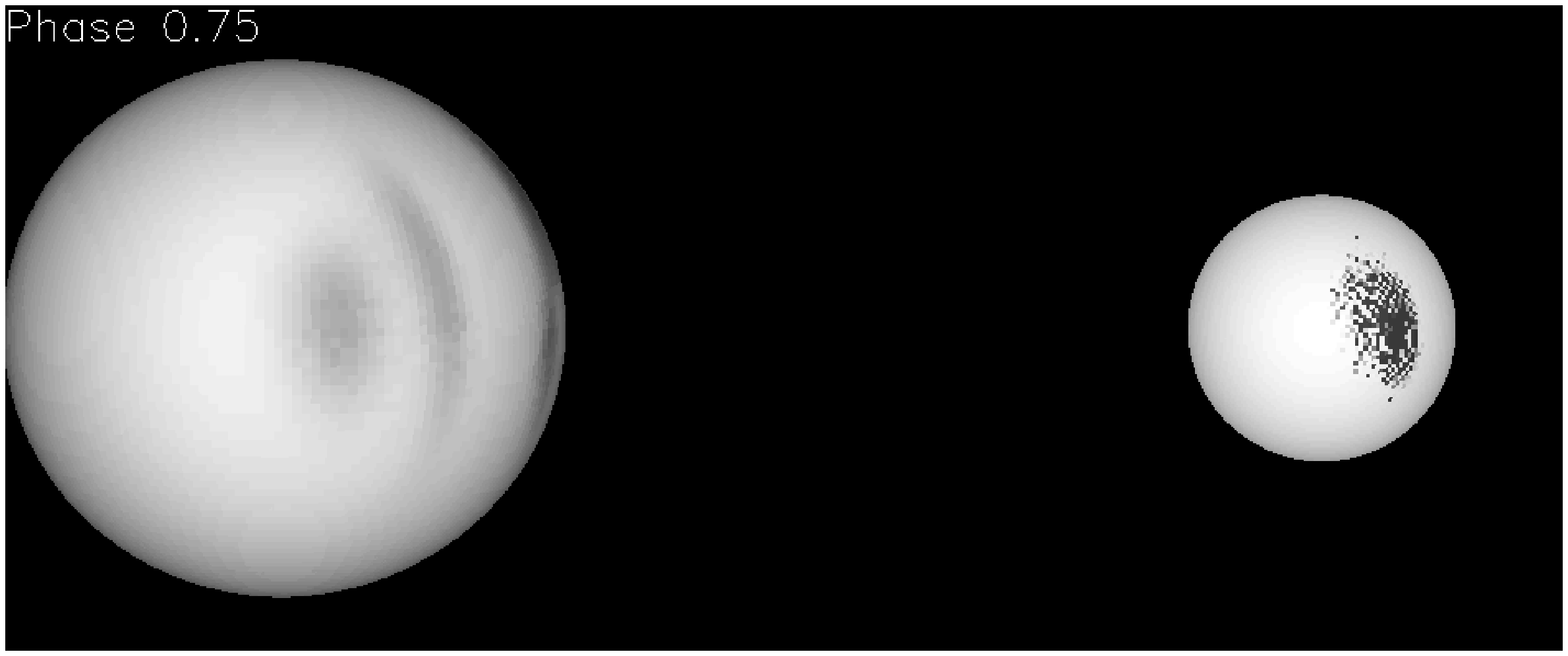}
\caption{Images of the binary system at orbital phases 0, 0.25, 0.5 and 0.75, in 2017 April. Due the very slow rotation speed of F star, the spot features on its surface probably are artefacts.}
\label{fig:i}
\end{figure*}

We also present the images of the binary system at orbital phases 0, 0.25, 0.5 and 0.75, of one orbital cycle in Fig. \ref{fig:i}, using the Doppler images of 2017 April. Since the rotational velocity of the F star is too slow (Table \ref{tab:par}), its reconstructed surface features are mainly related to the changes of the LSD line strength, and may not represent actual features on the F star's surface.

\begin{figure}
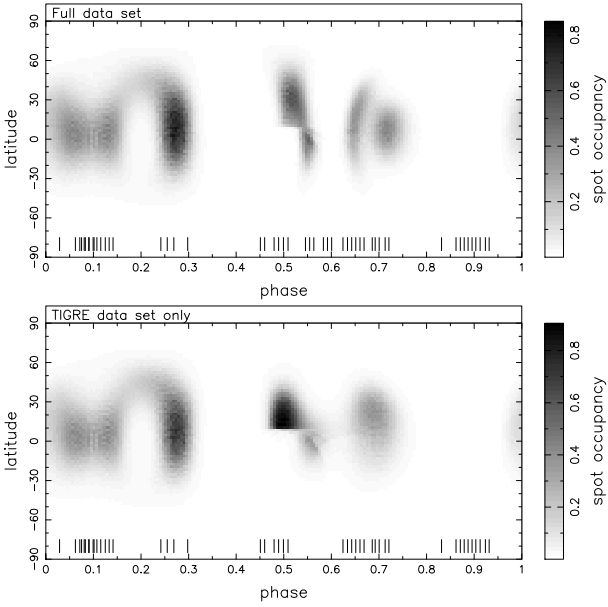

\centering
\includegraphics[bb = 71 19 359 597, angle=270, width=0.45\textwidth]{full.eps}
\includegraphics[bb = 71 19 359 597, angle=270, width=0.45\textwidth]{tigre.eps}
\caption{Comparison of the Doppler images derived from the combined TIGRE and Weihai data sets and the TIGRE data set only. The images are consistent where two data sets overlap, and the Weihai data set has a larger resolution.}
\label{fig:tigre}
\end{figure}

The Doppler image for 2017 April was derived from a combined data set collected by two telescopes, 1.2m TIGRE telescope and Weihai 1m telescope, at different observing sites in the same epoch to obtain a better phase sampling. However, the spectral resolution of two instruments differs by a factor of about 2.5, as described in Section 2. To show the effect of different spectral resolution on the image, we also derived a spot map from the data set acquired by the TIGRE telescope only, as shown in the lower panel in Fig. \ref{fig:tigre}. Since the Weihai data set only covered phases 0.5455--0.6006, we cannot use it to derive an image independently. As seen in Fig. \ref{fig:tigre}, the spot patterns of two images are very similar to each other, but the data set from Weihai 1m telescope helps to uncover finer spot structures, especially for phases 0.4--0.7. The spot image derived only from TIGRE data set shows an unresolved feature at phase 0.6, but the image from the combined spectra reveals that it is a connected structure of two smaller spots. One is a spot at phase 0.6 and another is a weaker feature at phase 0.7. This also hints that there should be more smaller spots on the surface of the K subgiant component of RS CVn, which are not resolved due to the limited spectral resolution.

\subsection{Surface differential rotation}

Starspots are the tracers of the stellar surface rotation, and the comparison between the Doppler images observed within several rotational cycles can reveal surface differential rotation patterns. The shear rate can be estimated either by the cross-correlation of two maps observed in close epoch \citep{don1997r} or by the shear imaging method which takes the differential rotation rate as a parameter directly in the Doppler imaging process \citep{petit2002}. Here we apply the cross-correlating map method to estimate the surface differential rotation rate for the K star of RS CVn.

\begin{figure}
\centering
\includegraphics[bb = 71 19 359 597, angle=270, width=0.85\linewidth]{dr1.eps}
\includegraphics[bb = 71 19 359 597, angle=270, width=0.85\linewidth]{dr2.eps}
\includegraphics[width=0.85\linewidth]{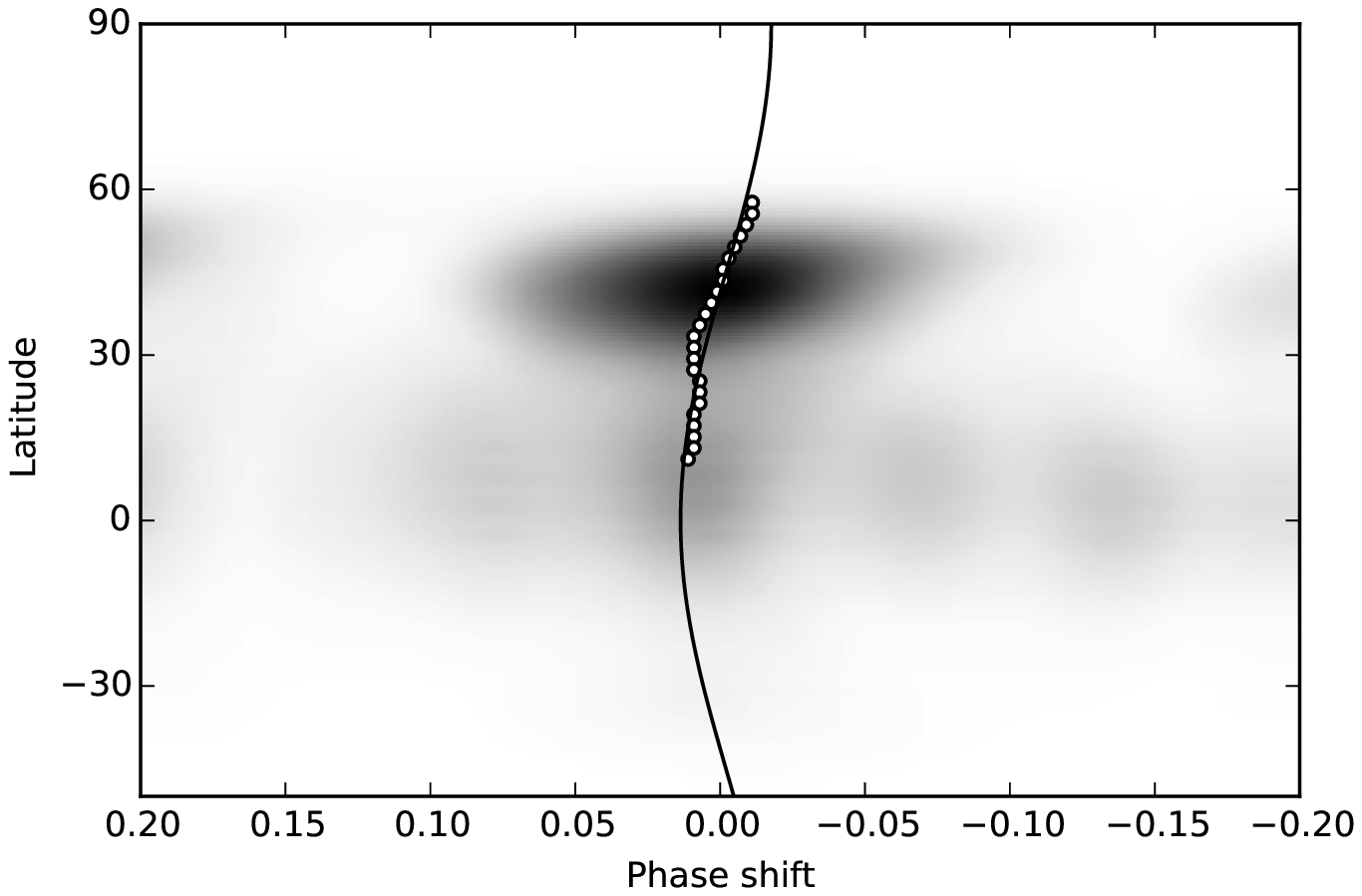}
\caption{The upper two panels show images derived from two consecutive rotation cycles. The bottom panel shows cross-correlating map for these two images. The maximum values of cross-correlation function at each latitude belt between latitude $0\degree$ and $60\degree$ are marked as open circles, and a curve is fitted with equation~(\ref{eq:dr}) given in Section 4.3. Note that the x-axis of cross-correlating map is inverse, since negative phase shift means shorter rotation period and faster rotation speed.}
\label{fig:dr}
\end{figure}

The spot configurations of two Doppler images will be biased if the phase coverage is incomplete and different. Hence we only chose the 2016 January data set to estimate the differential rotation rate, since it covered two consecutive rotation cycles with very similar phase coverage to each other. We partitioned the spectra of this data set to two sub sets covering the rotation cycles 0.59--1.24 and 1.63-2.27, respectively. Then we separately derived two Doppler images from these two sub sets, as shown in the upper two panels of Fig. \ref{fig:dr}. The cross-correlation function between two Doppler images, excluding the large phase gap of 0.30--0.55, was calculated latitude by latitude. The resulting cross-correlating map is shown as a grey-scale plot in the bottom panel of Fig. \ref{fig:dr}. Then we fit the peak phase shift of the cross-correlation function of each latitude belt between $10\degree$ and $60\degree$ (Fig. \ref{fig:dr}), since the high-latitude regions are featureless and the Doppler imaging is insensitive to the latitude of the equatorial spots. We assumed a solar-like, latitude-dependent surface rotation law, as
\begin{equation}
\Omega(l) = \Omega_{eq} - \Delta\Omega\sin^{2}l
\label{eq:dr}
\end{equation}
where $\Omega_{eq}$ is the angular velocity rate at the stellar equator, $l$ is the latitude and $\Delta\Omega$ is the difference between the rotation rates at the stellar equator and the pole. The results show that the surface rotation law for the K-type component of RS CVn follows $\Omega(l) = 1.293 (\pm 0.001) + 0.039 (\pm 0.003) \sin^{2}l\ [rad\ d^{-1}]$. However, due to the limited number of mid-to-high latitude spots, poor phase coverage and spot evolutions, the error is presumably underestimated.

\section{Discussion}

We have presented the maximum entropy reconstructed images of the active binary prototype RS CVn for 2004 February, 2016 January, 2017 April and 2017 November-December, derived from the high-resolution spectra collected with three telescopes at different observing sites. The K-type component of RS CVn exhibited starspot activity in all of four observing seasons. The surface images indicate complex spot patterns on the surface of the K star, which showed many small-to-moderate starspots, instead of one or two large active regions. The spot configurations revealed by our Doppler images are in excellent agreement with the results of \citet{eaton1993}. They found that 6--8 moderately sized spots on the surface of the K-type component are necessary for fitting the spectral line profiles of RS CVn. For the observed light curves, they also demonstrated that the multi-spot solution is better than the three-spot solution.

\begin{figure}
\centering
\includegraphics[width=0.75\linewidth]{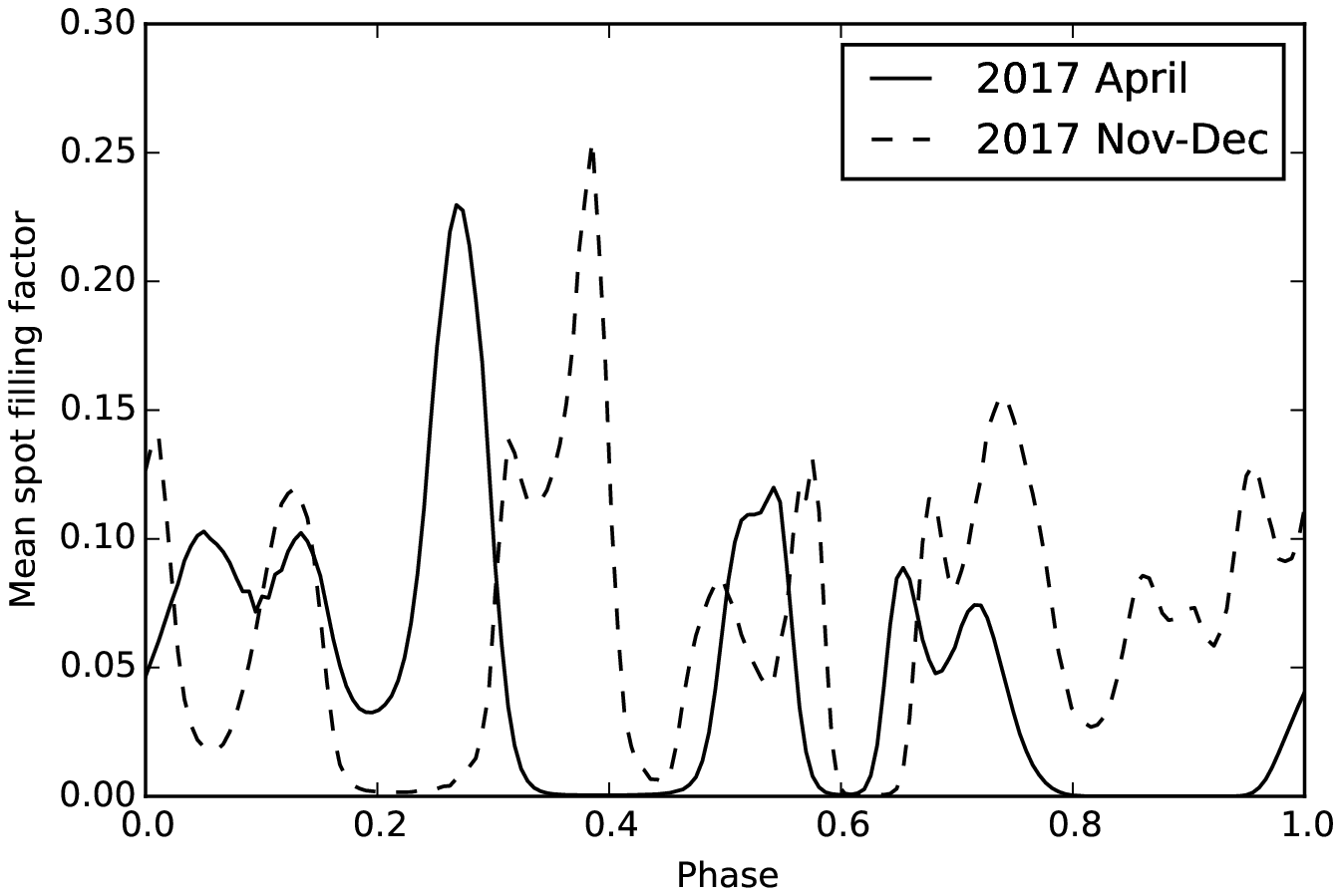}
\caption{Mean spot filling factor as a function of longitude, for 2017 April and 2017 November--December.}
\label{fig:fvl}
\end{figure}

The Doppler images also indicate a non-uniform longitudinal distribution of the starspots on the K star. In each observing season, we detected several active longitudes. We plot the mean spot filling factors as functions of longitude, for 2017 April and 2017 November--December images, in Fig \ref{fig:fvl}. The K star showed 4 active longitudes in 2017 April, but exhibited one more active region around phase 0.9 in 2017 November--December. The cross-correlation of two longitudinal distributions of mean spot filling factor between phases 0.1 and 0.8, where the common active longitudes existed, indicates a systematic phase shift of 0.08. Theoretically, the preferred longitudes may be produced by a non-axisymmetric dynamo \citep{moss2002}. For close binaries, the tidal force also plays an important role in the formation of active longitudes, since it can effectively affect the arising flux-tubes \citep{hol2003}. The active longitude migrations on the K star has been observed by many authors through long-term light curve modelling \citep{eaton1980,kang1989,hec1995}. They respectively revealed migration periods of 9.48yr, 9.4yr and 8.8yr for the active region. \citet{rod1995} detected up to three active longitudes on the K star of RS CVn and found their migration rates to be 0.1{\degree} per day in the direction of the rotation during 1963--1984 and 0.34{\degree} per day during 1988--1993. They attributed it as a result of the solar-like differential rotation on the surface of the K star. Comparing between our spot maps of K star in 2017 April and 2017 November--December, the systematic drift of 0.08 in phase, corresponding to -30{\degree} in longitude, may indicate a migration rate of 0.14{\degree} per day, but in the opposite direction of the stellar rotation, during the time interval of our two observing runs. But apparently we can not confirm this with the current data, since these two data sets were obtained 220 days apart and the spots may evolve much during the time span. \citet{eaton1979} found that the starspots on the K star have lifetime less than one year, but \citet{hec1995} observed starspot lifetime in a range between 2 and 6 yr.

Another notable feature is the latitude distribution of the starspots on the K star. Our Doppler images indicate that all spots were located below latitude 70{\degree}, and we did not find any high-latitude feature or polar cap in any of four observing seasons. The lack of polar spots revealed by our Doppler images is very consistent with previous studies. \citet{rod1995} demonstrated that no polar spot is required to fit their long-term light curves within the observation errors. \citet{eaton1993} revealed that there is no evidence for sizeable polar spots on the K star. From their observed line profiles, they even suspected whether it has polar starspots smaller than 18$\degree$ in radius.

The absence of high-latitude features on the K-type subgiant component of the prototype RS CVn is interesting, considering its relatively high rotational velocity (\vsini\ = 44.9 \kms). High-latitude spots and long-lived polar caps are commonly found on the surface of the rapidly-rotating, active component of RS CVn-type binary systems, revealed by both of photometric and spectroscopic studies. Several active single rapid rotators, such as AB Dor \citep{cam1994} and FK Com-type stars \citep{str1999}, were also reported to have large polar active regions. In our previous works, we found persistent high-latitude or polar features on the active binaries with various rotational speeds, such as II Peg (\vsini\ = 22 \kms; \citealt{xiang2014}), SZ Psc (\vsini\ = 67.7 \kms; \citealt{xiang2016}) and ER Vul (\vsini\ = 80 \kms; \citealt{xiang2015}).

Theoretical models suggested that the dominant Coriolis force and the meridional circulation can affect the magnetic flux transport within the convection zone to produce high-latitude magnetic emergence, and thus the latitude distribution of starspots is dependent on the rotational speed and the thickness of the convection zone \citep{sch1996,mac2004,hol2006}. Flux-tubes which emerge at low latitude can also be advected polewards by surface flows\citep{isik2007,isik2011}. For the lack of high-latitude spots on the Sun, \citet{sch1992} offered a simple scenario that the magnetic field at the bottom of the convection zone is ten times larger than the equipartition field strength, which results in the dominant buoyancy force. However, the tidal force and higher rotational speed make the dynamo process in close binaries more complex than that for single stars. An estimate from \citet{pat2002} show that the ratio of toroidal to poloidal magnetic field of RS CVn is less than 10 percent that of the Sun, which implies it is dominated by $\alpha^{2}$-$\Omega$ dynamo regime rather than $\alpha$-$\Omega$ one.

The surface differential rotation is an important factor in the stellar dynamo process. The cross-correlation of our Doppler images indicates an anti-solar surface shear rate of $\Delta\Omega = -0.039 \pm 0.003 ~rad~d^{-1}$ and $\alpha = \Delta\Omega/\Omega_{eq} = -0.030 \pm 0.002$ for the K star of RS CVn, which means the pole of the K subgiant rotates faster than the equator and laps it once every 161 d. The evidence of the surface latitudinal differential rotation of RS CVn was also found by \citet{rod1995}, but they inferred a solar-like shear rate of $\Delta\Omega = 2.3\degree \pm 1.6\degree d^{-1} (0.0401 \pm 0.0279 ~rad ~d^{-1}$) to explain the spot migrations found in the light curve analysis. Our estimate of the differential rotation rate was based on only two Doppler images. The phase coverage was not ideal and the value of shear rate is almost exclusively dependent on the large spot structure at phase 0.2. Meanwhile, the evolution of starspots also induce errors in the estimate of the differential rotation rate. More observations are required to confirm the differential rotation of the K-type star.

Recent Doppler imaging studies have detected the solar-like surface differential rotation on both of single and binary stars \citep{bar2000,dun2008,kri2014,ozd2016}, while some close binaries were reported to show the anti-solar differential rotations, which may be attributed to the tidal force \citep{kov2015,har2016}. \citet{gas2014, bru2017} showed that the different direction of the differential rotations are related to the predominance of the Coriolis force over the buoyancy force and vice versa. \citet{bar2005r} analyzed the differential rotation rates of 10 stars determined by Doppler imaging technique, and revealed that the shear rate is decreasing with the decrease of the effective temperature. \citet{kov2017} further investigated differential rotation of single and binary stars, and revealed that the trends of the surface shear rate of single stars are significantly different to those of close binaries, whose differential rotation is confined by the tidal force.

\section{Conclusion}

We have presented the first Doppler images of the active K-type subgiant component of the prototype RS CVn, derived from four data sets observed from 2004 to 2017. Based on the new reconstructed images, we summarise the results as follows.

1. The K-type component of RS CVn shows spot activity in all observing seasons. The Doppler images reveal complex spot patterns on the K star, which exhibits several small-to-moderate starspots.

2. The reconstructed images indicate that all spots are located below latitude 70{\degree}, and we do not find any high-latitude or polar spots.

3. The K star shows a non-uniform longitudinal spot distribution. We find several active longitudes on its surface and clues for spot migrations.

4. Using the cross-correlating technique, we derive an anti-solar differential rotation rate of $\Delta\Omega = -0.039 \pm 0.003 ~rad~d^{-1}$ and $\alpha = \Delta\Omega/\Omega_{eq} = -0.030 \pm 0.002$ for the K-type component of RS CVn. Due to the limited phase coverage and the evolution of spot patterns, the uncertainty of our differential rotation estimate is presumably higher.

In the future, more observations with shorter interval and longer time baseline are required to reveal the short-term evolution of starspots and the surface differential rotation on the K-type component of the active binary RS CVn.

\section*{Acknowledgements}
This work is supported by National Natural Science Foundation of China through grants No. 10373023, No. 11333006, No. 11603068. The joint research project between Yunnan Observatories and Hamburg Observatory is funded by Sino-German Center for Research Promotion (GZ1419). UW acknowledges funding by DLR, project 50OR1701. We would like to thank Prof. Jianyan Wei and Prof. Xiaojun Jiang for the allocation of observing time of the Xinglong 2.16-m telescope. We are very grateful to the anonymous referee for helpful comments and suggestions that significantly improved the clarity and quality of this paper. This work is based on data obtained with the TIGRE telescope, located at La Luz observatory, Mexico. TIGRE is a collaboration of the Hamburger Sternwarte, the Universities of Hamburg, Guanajuato and Li\`{e}ge. This work has made use of the VALD database, operated at Uppsala University, the Institute of Astronomy RAS in Moscow, and the University of Vienna.


\label{lastpage}
\end{document}